\documentclass[aps,prd,amsfonts,amssymb, amsmath, showpacs, showkeys, a4paper, nofootinbib, superscriptaddress, 11pt, raggedbottom]{revtex4-2}

\usepackage{braket}
\usepackage{bm}
\usepackage{graphicx}
\usepackage{slashed}
\usepackage{subfig}
\usepackage{mathrsfs}
\usepackage{color}
\usepackage{mathtools}
\usepackage{float}
\usepackage{wasysym}
\usepackage{siunitx}

\newcommand{\qbounce}{{\it{q}}{\sc{Bounce}}}				
\newcommand{\cannex}{{\sc{Cannex}}}				

\begin{document}


\title{\Large Screened Scalar Fields in the Laboratory and the Solar System}


\author{Hauke Fischer}
\affiliation{Technische Universit\"at Wien, Atominstitut, Stadionallee 2, 1020 Vienna, Austria}

\author{Christian K\"{a}ding}
\affiliation{Technische Universit\"at Wien, Atominstitut, Stadionallee 2, 1020 Vienna, Austria}

\author{Mario Pitschmann}
\email{mario.pitschmann@tuwien.ac.at}
\affiliation{Technische Universit\"at Wien, Atominstitut, Stadionallee 2, 1020 Vienna, Austria}


\begin{abstract}
The last few decades have provided abundant evidence for physics beyond the two standard models of particle physics and cosmology. As is now known, the by far largest part of our universe's matter/energy content lies in the `dark' and consists of dark energy and dark matter. Despite intensive efforts on the experimental as well as the theoretical side, the origins of both are still completely unknown. Screened scalar fields have been hypothesized as potential candidates for dark energy or dark matter. Among these, some of the most prominent models are the chameleon, symmetron, and environment-dependent dilaton. In this article, we present a summary containing the most recent experimental constraints on the parameters of these three models. For this, experimental results have been employed from the \qbounce{} collaboration, neutron interferometry, and Lunar Laser Ranging  (LLR), among others. In addition, constraints are forecast for the Casimir And Non Newtonian force EXperiment (\cannex{}). Combining these results with previous ones, this article collects the most up-to-date constraints on the three considered screened scalar field models.
\end{abstract}


\maketitle


\section{Introduction}
\label{sec:intro}
Dark energy (DE) and dark matter (DM) pose some of modern physics' greatest open problems. Modifications of general relativity (GR) in the form of scalar-tensor theories~\cite{Fujii2003} in which a scalar field is coupled to the gravitational metric tensor in the matter action are frequently employed in attempts to solve these conundrums~\cite{Clifton2011,Joyce2014}. While many of these theories lead to a universal coupling between the scalar field and the trace of the energy-momentum tensor of (Standard Model) matter and, consequently, a fifth fundamental force of Nature, such emergent effects are already tightly constrained within our Solar System~\cite{Dickey1994,Adelberger2003,Kapner2007}.

Screening mechanisms offer phenomenologically appealing ways for circumventing these constraints by rendering scalar fifth forces feeble in matter at least as dense as our Solar System. In turn, in environments that typically correspond to comparatively low matter densities, for example, on the edges of galaxies or galaxy clusters, screening mechanisms allow fifth forces to potentially even exceed gravity's strength. Consequently, scalar fifth forces could themselves give rise to observable phenomena at such scales and, for example, serve as potential alternatives to particle DM~\cite{Burrage:2016yjm,OHare:2018ayv,Burrage:2018zuj,Kading:2023hdb}. A zoo of scalar field models with screening mechanisms have been devised in the last decades. Among those so-called screened scalar fields, some of the most prominent ones are chameleons~\cite{Khoury2003,Khoury20032}, symmetrons~\cite{Dehnen1992, Gessner1992, Damour1994, Pietroni2005, Olive2008, Brax2010,Hinterbichler2010,Hinterbichler2011}, environment-dependent dilatons~\cite{Damour1994,Gasperini:2001pc,Damour:2002nv,Damour:2002mi,Brax:2010gi,Brax:2011ja,Brax2022} and galileons~\cite{Dvali2000,Nicolis2008,Ali2012}. In recent years, a large number of high-precision experiments and observations have been employed to constrain the parameter spaces of several of these screened scalar field models, see e.g., Refs.~\cite{Burrage:2016bwy,Burrage:2017qrf,Pokotilovski:2012xuk,Pokotilovski:2013tma,Jenke:2014yel,Baum:2014rka,Burrage:2014oza,Hamilton:2015zga,Lemmel:2015kwa,Burrage:2015lya,Elder:2016yxm,Ivanov:2016rfs,Burrage:2016rkv,Jaffe:2016fsh,Brax:2017hna,Sabulsky:2018jma,Brax:2018iyo,Cronenberg:2018qxf,Zhang:2019hos,ArguedasCuendis:2019fxj,Jenke:2020obe,Pitschmann:2020ejb,Brax:2021wcv,Qvarfort:2021zrl,Yin:2022geb,Betz:2022djh, Brax:2022olf, Hartley:2019wzu,Fischer:2023koa,Fischer:2023eww,Fischer:2024coj,Klimchitskaya:2024dvk,Rene}. In addition, first investigations of screened scalars within a quantum field theoretical framework have been carried out~\cite{Brax:2018grq,Burrage:2018pyg,Burrage:2019szw,Kading:2023mdk,Baez-Camargo:2024jia} and it has been suggested to study spacetimes with screened scalar-tensor theories in analogue gravity simulations~\cite{Hartley:2018lvm}.

In this article, we present an overview of existing experimental and observational constraints on chameleon, symmetron and environment-dependent dilaton models. For this, we add the most recent constraints on chameleons~\cite{Brax:2018grq,Yin:2022geb} and symmetrons~\cite{Brax:2022olf,Fischer:2023eww} to the summarised data from Ref.~\cite{Burrage:2017qrf}, and create plots combining constraints obtained from gravity resonance spectroscopy (\qbounce), Lunar Laser Ranging (LLR)~\cite{Fischer:2023koa} and neutron interferometry~\cite{Fischer:2023eww} for the environment-dependent dilaton. In addition, we present projected constraints for all three models expected to be obtained from the upcoming Casimir And Non Newtonian force EXperiment (\cannex{})~\cite{Sedmik:2021iaw,Rene}.

In addition to providing a brief review of existing constraints, this article updates the \qbounce{} constraints from Refs.~\cite{Jenke:2014yel,Burrage:2014oza,Cronenberg:2018qxf,Jenke:2020obe} for the symmetron and chameleon models. Previous analyses assumed an ideal vacuum density of $\rho_V=0$, exclusively used perturbation theory to compute the energy shifts that a screened scalar field would induce on a neutron bouncing in a gravitational field (which is a good approximation for the chameleon field \cite{Brax:2013cfa} but not for the symmetron field), and neglected parameter regions for which the symmetron field is in its symmetry-broken phase even inside the neutron mirror. This updated analysis employs the actual experimental residual gas density of $\rho_V=2.32\times 10^{-7}$ kg/m$^3$ in the vacuum chamber. For parameter regions where perturbation theory is inaccurate, a non-perturbative numerical treatment is used. Additionally, the symmetry-broken phase of the neutron is considered, as detailed in Ref.~\cite{Pitschmann:2020ejb}. Analogous to Ref.~\cite{Fischer:2023koa}, assumptions necessary for the analysis as, e.g., neglecting higher order couplings between scalar field and matter as well as neglecting the influence of the vacuum chamber on the scalar field have been properly considered. In order to physically justify these assumptions, all derived constraints have been cut off wherever this is required. As demonstrated in Appendix~\ref{AdditionalQBounce}, each of these improvements corrects the previous symmetron constraints by several orders of magnitude.

The article is organized as follows. In Sec.~\ref{sec:theory}, we provide the required theoretical background on the considered screened scalar field models, while in Sec.~\ref{sec:calc} our way of obtaining the constraints is described. Next, in Secs.~\ref{sec:chameleon}-\ref{sec:dilaton}, up-to-date constraints on the chameleon, the symmetron and the environment-dependent dilaton models are represented, respectively. Finally, in Sec.~\ref{sec:conclusion}, we draw our conclusions.


\section{Theoretical background}
\label{sec:theory}

In this article, we only discuss screened scalar field models with canonical kinetic terms in the Einstein frame in contrast to galileon models that require additional kinetic terms for implementing their screening via the Vainshtein mechanism~\cite{VAINSHTEIN1972393}. Consequently, all considered models for a scalar $\phi$ can be described by the following action
\begin{align}\label{eq:action}
S = \int  d^4x\, \sqrt{-g} \left( -\frac{m_\text{pl}^2}{2}\,R + \frac{1}{2}\,\partial_\mu\phi\,\partial^\mu\phi - V(\phi) \right) + \int d^4x\,\sqrt{-\tilde g}\,\mathcal{L}_\text{SM} (\tilde g_{\mu \nu},\psi_i)\>,
\end{align}
where $m_\text{pl}$ is the reduced Planck mass, $g_{\mu\nu}$ the Einstein frame metric, $\tilde{g}_{\mu\nu}$ the Jordan frame metric, $V(\phi)$ denotes the scalar's self-interaction potential characterising each individual screened scalar field model (see Tab.~\ref{tab:models}) and $\mathcal{L}_\text{SM}$ the Lagrangian describing the Standard Model (SM) fields $\psi_i$. The conformal transformation between Jordan and Einstein frame is given by $\tilde{g}_{\mu\nu} = A^2(\phi)g_{\mu\nu}$, where the form of the conformal factor $A(\phi)$ is determined by the particular scalar field model, see Tab.~\ref{tab:models}. In the Einstein frame, this leads to a universal coupling between scalar $\phi$ and the trace of the energy-momentum tensor $T^{\mu\nu}$. In turn, this coupling leads to a fifth force of Nature, which, for a test particle with mass $m$ in the non-relativistic limit is given by
\begin{align}
    \vec{f}_{\phi} &= - m\vec{\nabla}\text{ln}A(\phi) \nonumber\\
    &= - \beta (\phi) \frac{m}{m_{\text{pl}}} \vec{\nabla}\phi\>,
\end{align}
with the dimensionless coupling parameter
\begin{align}
    \beta(\phi) &:= m_{\text{pl}} \frac{d\ \text{ln}A}{d\phi}\>.
\end{align}
\begin{table}[H]
\centering
\renewcommand{\arraystretch}{1.5}
\begin{tabular}{|c||c|c|}
\hline\
Scalar Field & $V(\phi)$ & $A(\phi)$ \\
\hline\hline
\textit{Chameleon} & $\displaystyle\frac{\Lambda^{n+4}}{\phi^n}$ &$\displaystyle{\rm e}^{\phi / M_c}$ \\
\hline
\textit{Symmetron}  & $\displaystyle-\frac{\mu^2}{2}\,\phi^2 + \frac{\lambda_S}{4}\,\phi^4$ & $\displaystyle1 + \frac{\phi^2}{2M^2}$ \\
\hline
\textit{Dilaton} & $\displaystyle V_0\, {\rm e}^{-\lambda \phi /m_{\text{pl}}}$ &$\displaystyle1 + A_2\,\frac{\phi^2}{2m_{\text{pl}}^2}$ \\
\hline
\end{tabular}
\caption{\label{tab:models}The three considered scalar field models characterised by their particular forms of the potential $V(\phi)$ and the conformal factor $A(\phi)$. \textit{Chameleon:} the different chameleon models are distinguished by the parameter $n \in \mathbb{Z}^+ \cup 2\mathbb{Z}^-\setminus\{-2\}$. $\Lambda$ has the dimension of a mass and parameterizes the self-interaction of the chameleon. The mass scale $M_c$ describes the chameleon's coupling to matter. \textit{Symmetron:} $\mu$ is a tachyonic mass and $\lambda_S$ is a dimensionless self-coupling parameter. $M$ is a mass scale describing the symmetron coupling to matter. \textit{Dilaton:} The environment-dependent dilaton is characterised by a constant energy density $V_0$, a dimensionless self-coupling parameter $\lambda$, and a dimensionless constant $A_2$ parameterizing the coupling to matter.}
\end{table}
If the SM fields $\psi_i$ correspond to non-relativistic matter with density $\rho$, we can use $T^{\mu}_{\phantom{\mu}\mu} = \rho$. In this case, the scalar is subject to an effective potential \begin{align}\label{eq:effpot}
    V_{\text{eff}} (\phi; \rho)\>&:= V(\phi)+\rho A(\phi)\>,
\end{align}
which can be derived from Eq.~(\ref{eq:action}). Consequently, the equation of motion for the scalar $\phi$ is given by
\begin{align}\label{eq:eom}
    \Box \phi &= -V_{\text{eff}, \phi} (\phi; \rho)  \>.
\end{align}
Depending on the different possible choices for $V(\phi)$ and $A(\phi)$ as given in Tab.~\ref{tab:models}, the effective potential in Eq.~(\ref{eq:effpot}) gives rise to the chameleon mechanism~\cite{Khoury2003,Khoury20032} or the Damour-Polyakov mechanism~\cite{Damour1994}. While the former is characteristic for chameleon models, symmetrons are subject to the latter. Initially, the environment-dependent dilaton was also thought to be mainly screened by the Damour-Polyakov mechanism but in Ref.~\cite{Fischer:2023koa} it was demonstrated that for some parts of the dilaton parameter space the chameleon mechanism is actually dominant concerning the fifth force screening.


\section{Constraint calculation}
\label{sec:calc}

In this section, further details concerning the newly computed or re-analyzed constraints presented in this article are provided.  Wherever necessary, numerical algorithms have been employed to solve the scalar field equations of motion and, if applicable, the stationary Schrödinger equation. Refs.~\cite{Fischer:2024coj,Fischer:2023eww} provide further details concerning these algorithms.
Herein, new results or re-analyses are provided for the \qbounce{}, neutron interferometry and \cannex{} laboratory experiments as well as Lunar Laser Ranging. In the subsequent subsections, further details concerning the constraints from each of these experiments are presented. In order to justify the neglect of any higher order couplings, the analysis herein is restricted to
\begin{align}\label{eq:assumhighord}
    A(\phi) - 1 \ll 1 \>,
\end{align}
which requires excluding those regions of the constraint plots that do not fulfill this assumption. For \qbounce{} and \cannex{}, only interaction ranges in vacuum of up to $1$ mm are considered, in which case any influence from the vacuum chamber can be neglected. Furthermore, for \cannex{}, parameters for which the field does not decay to its potential minimum inside the upper mirror are excluded, while for neutron interferometry an interaction range of maximally $0.25$ mm within the cylindrical walls enclosing the chambers through which the superposed neutron beams traverse is considered. The latter is required to ensure that the minimum value $\phi_M$ of the scalar field within the walls can be used as a boundary condition.


\subsection{\qbounce{} constraints}

The \qbounce{} experiment is a realization of gravity resonance spectroscopy~\cite{Jenke:2014yel,PhysRevD.81.065019,Jenke:2011zz,Sponar:2020gfr}. It exploits the fact that ultracold neutrons, i.e. neutrons with very small kinetic energies, are totally reflected from most materials. In Earth's gravity, the quantum theoretical energy levels of a neutron are discrete and non-equidistant. This enables the realization of spectroscopy by inducing transitions between different energy levels. The very high precision enables effective searches for any hypothetical fifth forces. Specifically, measurements of the transition energies $\Delta E_{13}:= E_3-E_1$ and $\Delta E_{14}:= E_4-E_1$ have been carried out. The experimentally measured values are $ \Delta E_{13} =(1.9222\pm 0.0054)$ peV (expected value 1.9145 peV) and $\Delta E_{14} =(2.6874\pm 0.0074)$ peV (expected value 2.676 peV), consistent with the transition energies expected from a Newtonian gravitational potential alone~\cite{Cronenberg:2018qxf}. If it exists, a hypothetical scalar field would induce an additional potential. 

\qbounce{}'s experimental geometry, as realized in Ref.~\cite{Cronenberg:2018qxf}, can be approximated to an excellent degree by an infinitely extended single mirror, see e.g.~Refs.~\cite{Brax:2017hna, Pitschmann:2020ejb}, and the individual scalar field profiles computed by numerically solving the corresponding equations of motion~(\ref{eq:eom}). The transition energies are computed numerically by solving the stationary Schrödinger equation including Newtonian and scalar field effects
\begin{align}
    - \frac{1}{2 m_n} \frac{d^2 \Psi_n(z)}{dz^2}+U(z) \Psi_n(z)=E_n\Psi_n(z)\>,\label{SER}
\end{align}
where $m_n$ is the mass of the neutron, and the potential is given by 
\begin{align}
    U(z)&=m_ngz+U_X(z)\>,\nonumber\\
    U_X(z)&=\mathfrak{Q}_Xm_n\big(A_X(\phi)-1\big)\>,
\end{align}
with $X\in \{C, S, D\}$ (as abbreviations for chameleon, symmetron and dilaton, respectively) and the screening charge $\mathfrak{Q}_X$, which quantifies the amount of screening of the neutron, see Appendix~\ref{QFaktor}. Furthermore, the same relations as in Ref.~\cite{Fischer:2023eww} have been employed herein. 

Previous publications~\cite{Cronenberg:2018qxf,Jenke:2020obe,Fischer:2023koa} explored two distinct approximations for the coupling of neutrons to screened scalar fields, both treating the neutron as a classical sphere. In the fermi screening approximation, the neutron's radius is set to $R=0.5$ fm, a value grounded in QCD scattering data. Conversely, the micron screening approximation sets the neutron's radius at \SI{5.9}{\micro\meter} derived from its wavefunction's vertical extent.
In this investigation, we exclusively adhere to the fermi screening approximation. Notably, due to the absence of a well-defined micron screening approximation for neutron interferometry, such an approach becomes incomparable to other neutron experiments. Moreover, the wave function of neither experiment accurately conforms to a spherical model.
Additionally, fermi screening constraints are more conservative compared to micron screening constraints, resulting in notably weaker constraints. Consequently, the derived constraints may underestimate the true constraints. Further theoretical advancements are thus needed to determine the genuine coupling of neutrons to screened scalar fields.

Scalar field parameters are constrained, which would lead to a deviation of at least two standard deviations from the measured values, i.e. 
\begin{align}
    \Delta E_{13} &\not\in (1.9114,1.933)\text{ peV}\>,\nonumber\\
   &\text{or} \nonumber\\
   \Delta E_{14} &\not\in (2.6726,2.7022)\text{ peV}\>.\label{ConstraintCriterionq}
\end{align}
The current constraint criterion fails to account for correlations among $ \Delta E_{13}$, $ \Delta E_{14}$ and the scalar field parameters under scrutiny. Due to persisting theoretical uncertainties in energy shift predictions, notably attributable to the screening charge approximation, conducting a rigorous statistical analysis is presently unfeasible. However, it has been carefully checked that this simplified constraint criterion is a very good approximation to a more rigorous $\chi^2$-analysis, see Appendix~\ref{Constraintcriterieqbounce}, and is hence employed at the current level of theoretical description.


\subsection{Neutron interferometry constraints}

A neutron interferometer uses the wave nature of neutrons for performing interferometric measurements~\cite{Rauch74, RauchBook,Sponar:2020gfr}. Specifically, the path of a single propagating neutron is split into a superposition of two macroscopically separated beams before these paths are recombined and the neutron is finally detected. Each beam passes through a cylindrical chamber, one of which, denoted ``air'', is filled with air of ambient density to suppress the field, while the other, denoted ``vacuum'', actually contains helium with low pressure. In the latter, a scalar field typically takes on a larger field value than in the other chamber filled with air. If a scalar field indeed exists, the neutron would experience a different phase shift in each of the paths, which in the semi-classical limit is given by (see e.g. Ref.~\cite{Fischer:2023eww})
\begin{align}
    \delta\varphi_{X;P}(r) = - \frac{m_n}{k_0} \int\limits_{-L/2}^{L/2} U_{X;P}(r,z)\, dz\>,
\end{align}
with $X\in \{C, S, D\}$, $k_0$ being the wave number of the neutron, $r$ the radius from the center at which the neutron beam propagates through the chamber, $L$ the length of the chamber, $U_{X;P}(r,z)=\mathfrak{Q}_Xm_n\big(A_X(\phi)-A_X(\phi_{\text{Air}})\big)$  the scalar field potential in the chamber with pressure $P$, and the $z$-integration extending over the classical flight path (CFP) inside the corresponding chamber. 

Experimentally, the following two measurement modes have been used to constrain scalar field parameters:
\begin{enumerate}

\item{\textit{Profile mode}}

In this mode, the following phase shift is evaluated
\begin{align}
    \Delta\varphi_{X;P} = \delta\varphi_{X;P}(0) - \delta\varphi_{X;P}(0.015\text{ m})\>,
\end{align}
for both chambers, with the vacuum chamber having its pressure at $10^{-4}$ mbar, which corresponds to the lowest pressure measured in the experiment. The experiment actually measured 
\begin{align}
    \alpha:=\Delta\varphi_{X;\text{Vacuum}}-\Delta\varphi_{X;\text{Air}}<0\>.
\end{align}
This quantity is negative since the potential is more suppressed close to the chamber walls and due to $|\Delta\varphi_{X;\text{Air}}|<|\Delta\varphi_{X;\text{Vacuum}}|$. The experiment constrains
\begin{align}
    \alpha < -3.55^{\circ}\>,
\end{align}
as will be further detailed in Appendix~\ref{ProfileModeConstraints}. 

\item{\textit{Pressure mode}}

In this mode, the following quantity is measured instead
\begin{align}
    \gamma:=\delta\varphi_{X;P_0}(0) - \delta\varphi_{X;P_1}(0)<0\>,
\end{align}
with a vacuum pressure of $P_0=2\times 10^{-4}$ mbar and a reference pressure of $P_1= 10^{-2}$ mbar. Scalar field parameters are constrained if
\begin{align}
    \gamma < -5.44^{\circ}\>,
\end{align}
as is detailed in Appendix~\ref{PressureModeConstraints}.
\end{enumerate}

Further details on the specific experimental setup used to obtain the constraints in the present article can be found in Ref.~\cite{Fischer:2023eww}.

\subsection{Computing observables for Lunar Laser Ranging}

Lunar Laser Ranging (LLR) allows for the measurement of the Moon's orbit with high precision and consequently tests general relativity as well as potential deviations from it, for example, due to the presence of a scalar fifth force~\cite{muller2019lunar}. For this, the Earth-Moon distance is measured by firing a laser beam at the retroreflectors that were installed on the Moon's surface during the Apollo missions and detecting the reflection back on Earth. The propagation time of the laser pulse during its travel from Earth to Moon and back provides the corresponding distance to high precision. For previous investigations employing LLR in the context of symmetron and chameleon fields, see Refs.~\cite{Sakstein:2017pqi,Kraiselburd:2015vyf,Zhang:2019hos}. 

The measured value for violations of the equivalence principle is~\cite{Hofmann:2018myc}
\begin{align}
\delta_{\text{em}}\simeq \frac{a_{\phi \earth}-a_{\phi \leftmoon}}{a_G} = (- 3 \pm 5) \times 10^{-14}\>,
\end{align}
where $a_{\phi \earth}$ and $a_{\phi \leftmoon}$ refer to the scalar field induced accelerations of the Earth and Moon towards the Sun, whereas $a_G$ is the regular Newtonian acceleration towards the Sun.
For the constraint plots in this article, parameters are constrained for which the scalar field contribution leads to at least a $2\sigma$-deviation of the measured value corresponding to
\begin{align}
    \delta_{\text{em}} \not \in (-1.3\times 10^{-13},7\times 10^{-14})\>.
\end{align}
The perihelion precession of the Moon has been computed, e.g. in Ref.~\cite{MarioHabil}, and leads to parameter constraints given by \cite{Adelberger:2003zx}
\begin{align}
\left|\frac{\delta \Omega}{\Omega}\right| \simeq \frac{R^2}{G M_{\earth}}\left|\delta f(R) + \frac{R}{2}\, \delta f'(R)\right| \leq 1.6 \times 10^{-11}\>, \label{C2}
\end{align}
where $R$ is the maximum Earth-Moon separation and use has been made of 
\begin{align}
    \delta_{\text{em}} &\simeq  -\frac{(\mathfrak{Q}_{\earth}-\mathfrak{Q}_{\leftmoon})}{a_G}\frac{dA(\phi)}{dr}\bigg|_{r=1\,\text{AU}}\>,\nonumber\\
    \delta f(r) &\simeq  \mathfrak{Q}_{\leftmoon}\frac{dA(\phi)}{dr}\>.
\end{align}


\subsection{Computing the pressure in the \cannex{} experiment}

Among the high-precision experiments is the Casimir And Non-Newtonian force EXperiment (\cannex), which has successfully completed its proof-of-principle phase and will soon commence operation~\cite{Sedmik:2021iaw, Rene}. It is the first experiment designed to measure the Casimir force employing two plane parallel plates. Due to this special geometry, interfacial as well as gravity-like forces are maximized leading to increased sensitivity. This experiment allows to probe a wide range of dark sector forces, Casimir forces in and out of thermal equilibrium as well as gravity. 

\cannex{} measures the pressure 
\begin{align}
P(d) = \frac{\rho_M}{\rho_M-\rho_V}\,\big(V_{\text{eff}}(\phi_V,\rho_V)-V_{\text{eff}}(\phi_0(d),\rho_V)\big),
\end{align}
 where $\rho_M$ and $\rho_V$ are the densities of the plates and of the surrounding vacuum, $d$ is the distance between both plates and $\phi_0(d)$ is the value of the scalar field in the middle between both plates, as well as the pressure gradient
\begin{align}
\partial_d P(d) \simeq \frac{P(d+\delta)-P(d-\delta)}{2\delta}.
\end{align}
The latter approximation holds for small enough $\delta$. The sensitivity of these measurements at one standard deviation $(1\sigma)$ have already been analyzed in detail as function of $d$ in Ref.~\cite{Sedmik:2021iaw}. Herein, parameter constraints are derived for
\begin{align}
    |P(d)|>2\sigma(d)\>, \quad \text{or} \quad |\partial_d P(d)|&> 2\sigma(d)\>.
\end{align}
For the derivation of prospective constraints, it has been used that the plate separation and vacuum density can be varied between $\SI{3}{\micro\meter} < d < \SI{30}{\micro\meter}$ and  $5.3\times 10^{-12}\ \text{kg}/\text{m}^3 < \rho_V < 2.6\ \text{kg}/\text{m}^3$, respectively. More details about this experiment can be found in Ref.~\cite{Fischer:2023koa}.


\section{Results}
This section provides succinct discussions of the obtained constraints for the chameleon, symmetron and dilaton model.
\subsection{Chameleon constraints}
\label{sec:chameleon}

\begin{figure}[H]
\includegraphics[width=\linewidth]
{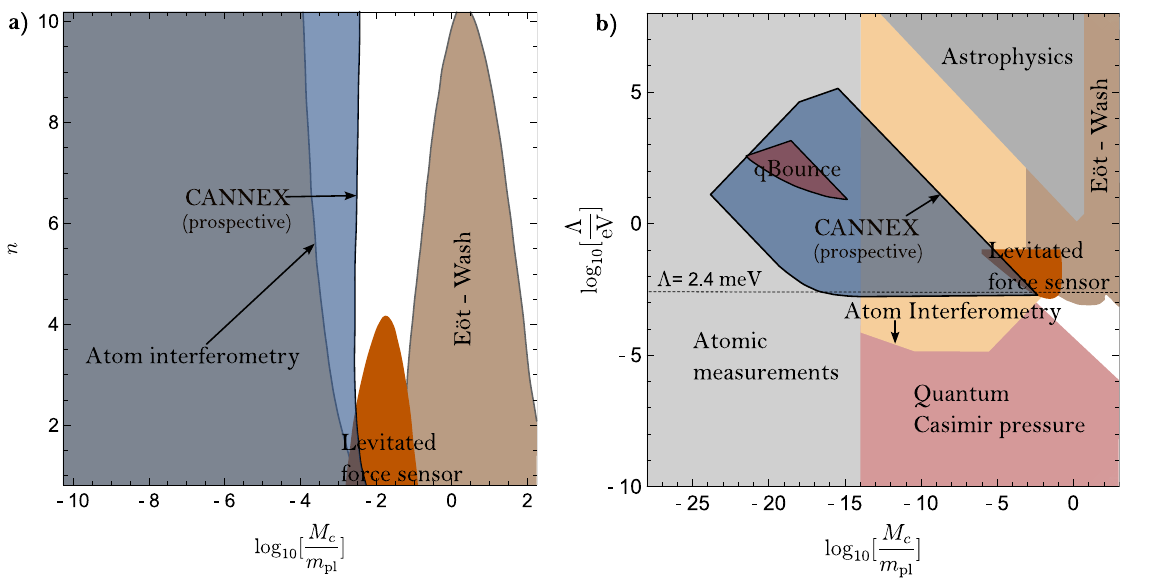}
\caption{Constraints on the parameter space of chameleon models based on the review~\cite{Burrage:2017qrf} and including results from quantum Casimir pressure~\cite{Brax:2018grq}, levitated force sensor measurements~\cite{Yin:2022geb}, and, for \qbounce{} and \cannex{} (the last only prospective), the analysis made in this article. 
\textbf{a)} Here, the parameter $\Lambda$ has been fixed to the DE scale of $2.4$ meV. The blue area shows the combined prospective constraints of pressure and pressure gradient measurements on chameleon interactions resulting from \cannex{}. \qbounce{} and neutron interferometry can set no constraints. This figure was adapted from Ref.~\cite{Rene}. \textbf{b)} The blue area shows the prospective \cannex{} constraints for the chameleon model with $n=1$, which are expected to fully overlap with already existing constraints. Constraints from \qbounce{} are depicted in the small dark red area that overlaps with atomic measurements and \cannex{}. The strong discrepancy of the computed \qbounce{} constraints in this work and previous constraints obtained in Refs.~\cite{Jenke:2014yel,Burrage:2014oza} is further elucidated in Appendix~\ref{QFaktor}.
\label{fig:CHAMELEON1}}
\end{figure}

In Fig.~\ref{fig:CHAMELEON1}, all currently available chameleon constraints on the parameters $n$ and $M_c$ for $\Lambda$ being set to the DE scale of $2.4$ meV are shown as well as those on the parameters $\Lambda$ and $M_c$ for the case $n=1$. 

For this, the analytically exact solutions from Ref.~\cite{Ivanov:2016rfs} are used to compute the prospective \cannex{} constraints, while the one-mirror solution for \qbounce{} is computed numerically.

Our analysis shows that \qbounce{} has set `weak' constraints for $n=1$ but no constraints for $\Lambda=2.4$ meV and small positive $n$. The large discrepancy to previous chameleon constraints published in Refs.~\cite{Jenke:2014yel,Burrage:2014oza} are primarily rooted in a different definition of the screening charge, further elaborated on in Appendix~\ref{QFaktor}.

For \cannex{}, our analysis predicts constraints both for $n=1$ as well as for other small values of $n$ if $\Lambda=2.4$ meV. However, constraints on parts of the parameter space that were not previously constrained by other experiments can only be obtained in the latter case.
Neutron interferometry is currently not capable of producing new constraints for the cases considered in Fig.~\ref{fig:CHAMELEON1}. A static chameleon field always fulfills the inequality
\begin{align}
    \phi_M \leq \phi(x) \leq \phi_V\>.
\end{align}
The quantity 
\begin{align}
   \delta_\text{max}\varphi_C&:= \mathfrak{Q}_C\,\frac{m_n^2L}{k_0M_c}\,\text{max}\{\left(|\phi_V-\phi_{\text{Air}}|\right),\left(|\phi_M-\phi_{\text{Air}}|\right)\}\>,
\end{align}
where $L=9.4$ cm, can be computed analytically and used to bound the largest expected phase shift. It should be noted that
\begin{align}
    &\big|\alpha\big| 
    \leq  4\delta_\text{max}\varphi_C\>,
\end{align}
as well as 
\begin{align}
     |\gamma|
     \leq   2\delta_\text{max}\varphi_C\>.
\end{align}
Over all unconstrained parts of the parameter space one has
\begin{align}
     4\delta_\text{max}\varphi_C \ll 1^{\circ}\>.
\end{align}
Consequently, no new constraints can be obtained from neutron interferometry. Obtained constraints would be many orders of magnitude `weaker' than existing constraints, rendering a more detailed analysis for this experiment futile in the case of chameleons. 


\subsection{Symmetron constraints}

\begin{figure}[H]
\includegraphics[width=\linewidth]
{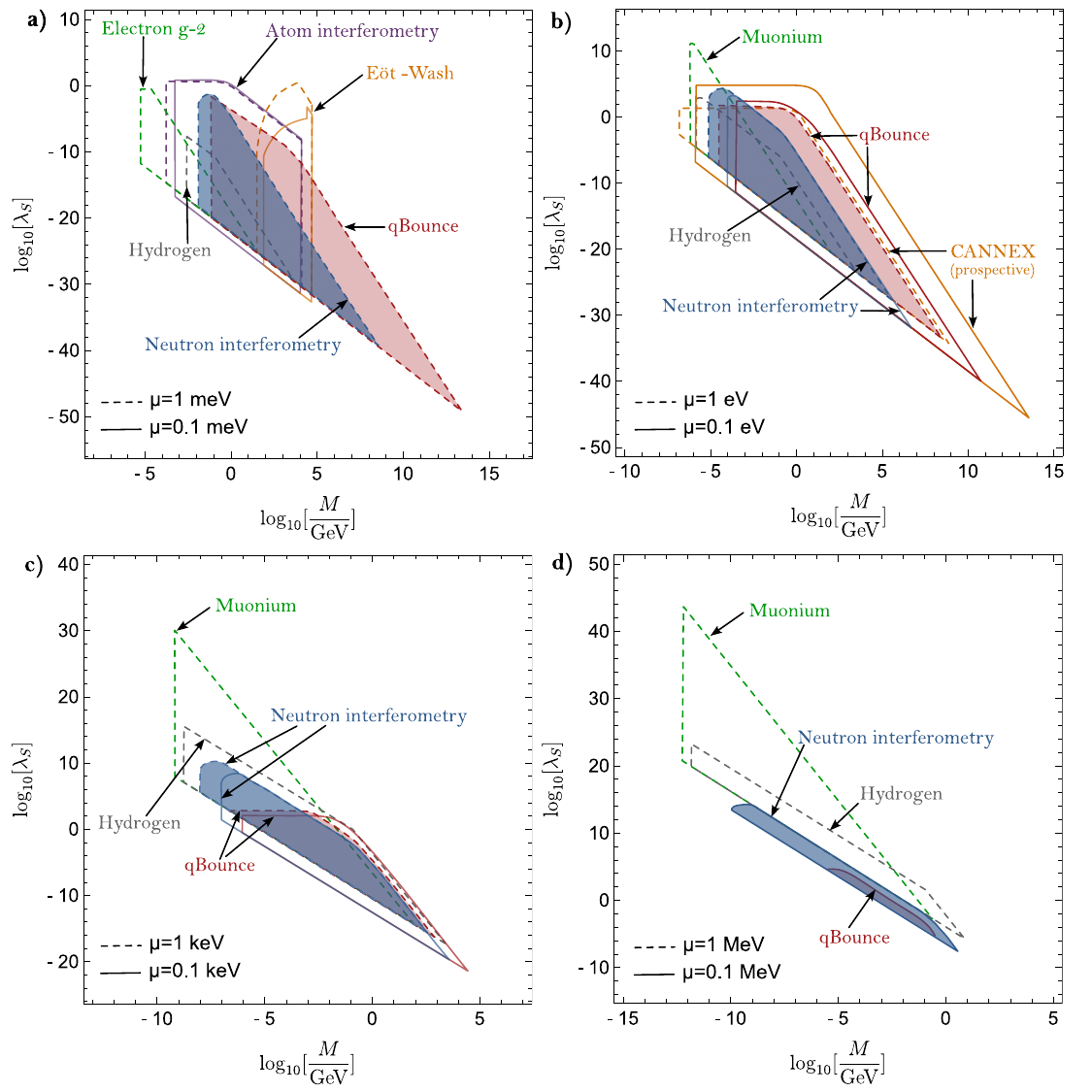}
\caption{Constraints on the parameter space of the symmetron model for different values of the tachyonic mass $\mu$ are depicted based on the review~\cite{Burrage:2017qrf}, containing Eöt-Wash results~\cite{Upadhye:2012rc} and the atom interferometry analysis from Ref.~\cite{Jaffe:2016fsh}; investigations of hydrogen, muonium and the electron (g-2)~\cite{Brax:2022olf}; and our new analysis for \qbounce{}, neutron interferometry and \cannex{} (the last only prospective).
\textbf{a)} Most table-top experiments are sensitive to the $\mu$ values displayed here since the symmetron range is close to $1$ mm.  \textbf{b)} \cannex{} is expected to lead to dominant constraints since the range is approximately \SI{1}{\micro\meter}, which is close to the plate separation. Eöt-Wash and atom interferometry are currently not able to set constraints for these values of $\mu$. \textbf{c)} Only some quantum experiments can still set constraints for these interaction ranges. \textbf{d)} For these values of $\mu$, the screening of nucleons is strong enough that they reach their limit for probing symmetrons, which affects the \qbounce{} and hydrogen constraints. In contrast, Ref.~\cite{Brax:2022olf} argues that muonium is always unscreened, due to its composition of effectively point-like particles, and can hence still set substantial constraints.
\label{fig:SYMMETRON1}}
\end{figure}

All existing constraints on the symmetron model derived from tabletop experiments are depicted in Fig.~\ref{fig:SYMMETRON1}. Notably, astrophysical experiments are unable to impose constraints on the $\mu$ values discussed in this article.
 
For the re-analysis of the \qbounce{} constraints originally presented in Ref.~\cite{Jenke:2020obe,Cronenberg:2018qxf}, the analytically exact solutions of Refs.~\cite{Brax:2017hna,Pitschmann:2020ejb} are employed. However, for the prospective \cannex{} constraints, the two mirror solution is computed by solving the differential equation numerically. The constraints are based on the most recent analysis of \cannex{} in Ref.~\cite{Rene}. Constraints from neutron interferometry are based on  Ref.~\cite{Fischer:2023eww}.  In Appendix~\ref{NewConstraints}, we delineate how the newly computed symmetron constraints compare to those previously derived in Ref.~\cite{Cronenberg:2018qxf}, elucidating the impact of each innovation on the resulting constraints.


\subsection{Dilaton constraints}
\label{sec:dilaton}

\begin{figure}[H]
\includegraphics[width=\linewidth]
{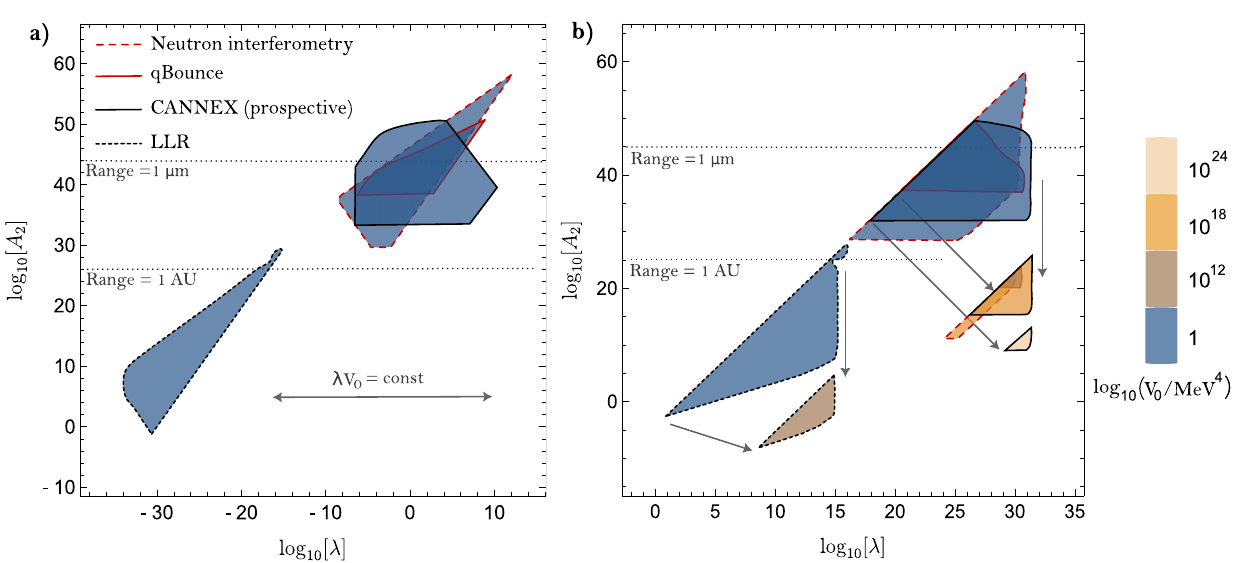}
\caption{Re-computed combined constraints from Refs.~\cite{Fischer:2023eww,Fischer:2023koa} on the parameters of the environment-dependent dilaton are presented. The interaction range of the dilaton with $\log_{10} (V_0 / {\rm MeV}^4) = 1$ is plotted for two different vacuum densities, the 1 \si{\micro\metre} and 1 AU contours correspond to  $\rho_V = 2.32 \times 10^{-7}$ kg/m$^3$ (\qbounce) and $\rho_V = 1.67 \times 10^{-20}$ kg/m$^3$ (interplanetary medium), respectively. \textbf{a)} The constraint areas shift towards lower values of $\lambda$ as $V_0$ increases without altering their shape. This phenomenon occurs because the fifth force effectively relies on the product of $V_0$ and $\lambda$ rather than their individual values in this regime. \textbf{b)} The constraint areas exhibit systematic shifts and cuts as illustrated by the indicated arrows. This figure has been adapted from Ref.~\cite{Fischer:2023koa}. 
\label{fig:DILATON}}
\end{figure}

All existing constraints on the environment-dependent dilaton model were originally obtained in Refs.~\cite{Fischer:2023eww,Fischer:2023koa} and are based on experimental results from \qbounce{}, LLR, and neutron interferometry. Prospective constraints for \cannex{} were also presented in Ref.~\cite{Fischer:2023koa}. A combined summary of these constraints can be found in Fig.~\ref{fig:DILATON}.


\section{Conclusions}
\label{sec:conclusion}

In this article, the most recent experimental constraints on the parameter spaces of some of the most popular screened scalar field models are summarized, i.e., the chameleon, symmetron and environment-dependent dilaton. Among others, experimental results from the \qbounce{} collaboration, neutron interferometry, and LLR have been employed. In addition, future constraints for \cannex{} are predicted. The results herein collect the most up-to-date constraints on the considered screened scalar field models. Furthermore, several improvements to previously obtained constraints have been made. In some cases, this led to deviations by several orders of magnitude compared to previously obtained results.


\begin{acknowledgments}
The authors are grateful to H.~Abele, T.~Jenke, and R.~I.~P.~Sedmik for useful discussions and comments on the manuscript. This article was supported by the Austrian Science Fund (FWF): P 34240-N, and is based upon work from COST Action COSMIC WISPers CA21106, supported by COST (European Cooperation in Science and Technology).
\end{acknowledgments}


\appendix
\appendix
\label{App:AQI}


\section{Additional information on \qbounce{} constraints} \label{AdditionalQBounce}
In this section, a detailed analysis concerning all improvements in deriving symmetron constraints with respect to earlier investigations is provided as well as confirmations that a $\chi^2$-data analysis at the current level of theoretical description can safely be neglected.


\subsection{Comparison of previous and new \qbounce{} analysis}\label{NewConstraints}

\begin{figure}[H]
\begin{center}
\includegraphics[width=\linewidth]
{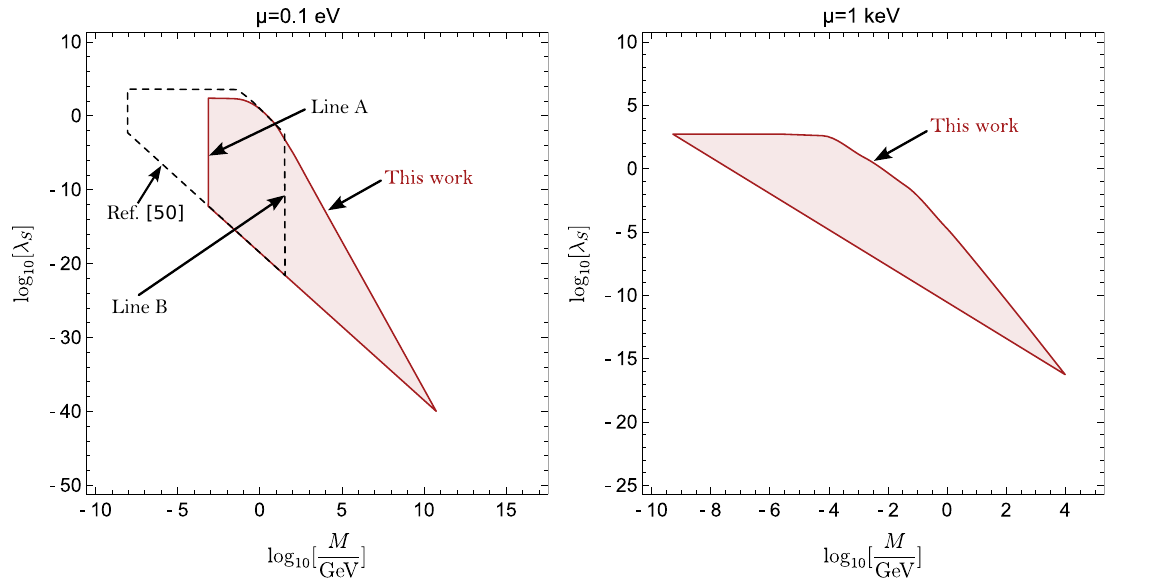}
\caption{Comparison between \qbounce{} constraints obtained in Ref.~\cite{Jenke:2020obe} and the constraints computed herein. \textbf{a)} The symmetron parameter $\mu$ has been set to 0.1 eV. Left of \textit{line A}, no symmetron solutions exist, since the symmetron is in its symmetric phase inside the vacuum as well as mirror regions. Right of \textit{line B}, the symmetron is in its symmetry-broken phase inside the mirror. \textbf{b)} Here, $\mu$ has been set to 1 keV, a value for which no constraints have been obtained in Ref.~\cite{Jenke:2020obe}.}
\label{fig:NewVsOldSymm}
\end{center}
\end{figure}

Fig.~\ref{fig:NewVsOldSymm} demonstrates a substantial deviation between the previous analyses from Ref.~\cite{Jenke:2020obe} and the new enhanced analyses provided herein. Each enhancement, such as considering the real finite vacuum density, parameter regions for which the symmetron is in its symmetry-broken phase inside the mirror and employing non-perturbative methodologies for energy computation, rectifies the analysis significantly.

It is worth noting that according to perturbation theory, \qbounce{} struggles to effectively constrain the symmetron field for $\mu$-values exceeding 10 eV. However, a numerical solution of the Schrödinger equation reveals significant constraints achievable up to $\mu = 0.1$ MeV. This notable discrepancy arises because in perturbation theory energy shifts are calculated by employing the unperturbed wave function, which extends approximately \SI{100}{\micro\meter}, according to
\begin{align}
\delta E_{pq} = \int_{-\infty}^{\infty}dz\,U_X(z)\left(\left|\Psi^{(0)}_p(z)\right|^2-\left|\Psi^{(0)}_q(z)\right|^2\right).
\end{align}
At $\mu = 10$ eV, the vacuum interaction range of the symmetron field is approximately \SI{0.1}{\micro\meter}. Consequently, the field quickly reaches its vacuum expectation value (VEV), causing the unperturbed wave function to perceive an almost constant potential shift, aside from the region \SI{0.1}{\micro\meter} immediately above the mirror, which contributes only minimally to the integral. This results approximately in
\begin{align}
\delta E_{pq} &\simeq U_X^{VEV}\int_{-\infty}^{\infty}dz\left(\left|\Psi^{(0)}_p(z)\right|^2-\left|\Psi^{(0)}_q(z)\right|^2\right) \nonumber\\
&= 0\>,
\end{align}
where $U_X^{VEV}$ is the value of $U_X(z)$ corresponding to the VEV. However, strong scalar fields at such short ranges distort the neutron wave functions, resulting in significantly different energy shifts for various energy states, which would be readily observable in experiments. The numerical solution of the Schrödinger equation provided herein accounts for that and hence allows to extend previous constraints. 

Thus, each improvement in the present analysis significantly changes existing constraints. In contrast, a full statistical data analysis to obtain constraints can safely be neglected at the current level of theoretical description, as will be demonstrated in the following section.


\subsection{Constraint criteria for \qbounce{}}\label{Constraintcriterieqbounce}

\begin{figure}[H]
\centering
\includegraphics[width=0.6\linewidth]
{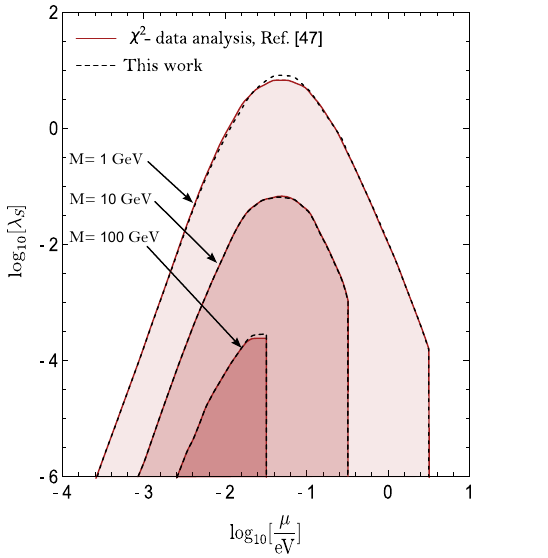}
\caption{The shaded regions illustrate parameter constraints (employing the fermi screening approximation) derived in Ref.~\cite{Cronenberg:2018qxf} for various fixed values of $M$. Dotted lines mark the boundary of the constraint regions computed with the simplified constraint criterion outlined in Eq.~(\ref{ConstraintCriterionq}). The constraints depicted in this plot were computed using previously made theoretical assumptions, as is detailed in Appendices~\ref{NewConstraints} and \ref{Constraintcriterieqbounce}. Updated constraints using the most recent theoretical assumptions are depicted in Fig.~\ref{fig:SYMMETRON1}.
\label{fig:Constraints}}
\end{figure}

In Ref.~\cite{Cronenberg:2018qxf}, constraints on the symmetron model from \qbounce{} have been established through a comprehensive $\chi^2$-data analysis, considering correlations between $\Delta E_{13}$, $\Delta E_{14}$ as well as each symmetron parameter $\mu$, $\lambda_S$ and $M$. However, this analysis is based on different theoretical assumptions, including a vacuum density of $\rho_V=0$, exclusive utilization of perturbation theory for computing energy shifts, neglecting the symmetry-broken phase of the symmetron inside the neutron mirror and employing different cutoff criteria. Here, we replicated this earlier investigation under the previously made assumptions but employed the constraint criterion outlined in Eq.~(\ref{ConstraintCriterionq}) instead of a full $\chi^2$-data analysis. This allows to assess the error incurred by forgoing a comprehensive statistical examination. As demonstrated in Fig. \ref{fig:Constraints}, this error is negligible in a log-log plot, validating our adoption of a simpler constraint criterion.


\section{Constraint criteria for neutron interferometry}

In this section, further details on the constraint criteria for neutron interferometry are provided.


\subsection{Profile mode} \label{ProfileModeConstraints}

The experimentally measured value is $\alpha=(0.56\pm 2.50)^{\circ}$. To establish a 95\% confidence level, we assume that the error follows a normal distribution and acknowledge that $\alpha<0$. Accordingly, we model the normal distribution with a mean of 0.56$^{\circ}$ and a standard deviation of 2.50$^{\circ}$. Given the predicted negativity of $\alpha$, our objective is to determine $x$ such that the probability of measuring $\alpha$ within the interval $(-\infty,x)$ is only 5\%, allowing us to  exclude $\alpha$-values in this interval with a 95\% CL. Hence, $x$ is obtained by solving 
\begin{align}
    \frac{1}{2.50^{\circ} \sqrt{2\pi}}\int_{-\infty}^{x}e^{-\frac{1}{2}\big(\frac{s-0.56^{\circ}}{2.50^{\circ}}\big)^2}ds=0.05\>,
\end{align}
which yields $x\simeq-3.55^{\circ}$, allowing us to constrain scalar field parameters for which 
\begin{align}
    \alpha < -3.55^{\circ}\>.
\end{align}


\subsection{Pressure mode}\label{PressureModeConstraints}

The experimentally measured result is $\gamma=(0.37 \pm 3.53)^{\circ}$. Taking into account that $\gamma<0$ and assuming that the error is normally distributed, $y$ is computed such that
\begin{align}
    \frac{1}{3.53^{\circ} \sqrt{2\pi}}\int_{-\infty}^{y}e^{-\frac{1}{2}\big(\frac{s-0.37^{\circ}}{3.53^{\circ}}\big)^2}ds=0.05\>,
\end{align}
providing the result $y\simeq - 5.44^{\circ}$. Hence, we constrain scalar field parameters for which 
\begin{align}
    \gamma < -5.44^{\circ}\>.
\end{align}


\section{Screening charge} \label{QFaktor}

In this work, we adopt the following definition of the screening charge for the dilaton and chameleon field, as derived in Refs.~\cite{Brax2022,Fischer:2023eww}
\begin{align}
    \mathfrak{Q} := \frac{3}{\mu^2_M R^2} \frac{1-\frac{1}{\mu_M R}\, \text{tanh}(\mu_M R)}{1+\frac{\mu_V}{\mu_M}\,\text{tanh}(\mu_M R)}\>,
\end{align}
where $\mu_M$ represents the chameleon or dilaton mass, respectively, within a neutron, $\mu_V$ denotes the corresponding mass within the surrounding vacuum and $R$ is the radius of the sphere. 
This definition ensures that 
\begin{align}
  \mathfrak{Q}\rightarrow\begin{dcases}
    0\>, & \text{for ``screened" bodies with }\mu_M R \rightarrow \infty\>,\\
    1\>, & \text{for ``unscreened" bodies with }\mu_M R \rightarrow 0\>.
  \end{dcases}
\end{align}
In contrast, Ref.~\cite{Burrage:2014oza} calculated \qbounce{} chameleon constraints using an alternative definition of the screening charge, sometimes treating the neutron as a test particle even when $\mu_M R\gg 1$, leading to significantly larger constraints. Ref.~\cite{Jenke:2014yel}, on the other hand, has neglected neutron screening altogether. This disparity underscores the imperative for further theoretical advancements to transcend heuristic approximations and accurately determine the true coupling of neutrons to the individual scalar field. The analysis of the symmetron field herein uses the screening charge derived in Refs.~\cite{Pitschmann:2020ejb,Brax:2017hna}, which has already been employed in the original analyses in Refs.~\cite{Cronenberg:2018qxf,Jenke:2020obe}. For some further background information concerning the methodology of screening charges, see e.g. Appendix~A of Ref.~\cite{Brax:2017hna}.


\bibliography{refs}

\providecommand{\href}[2]{#2}\begingroup\raggedright\begin{thebibliography}{10}

\bibitem{Fujii2003}
{Fujii, Yasunori and Maeda, Kei-ichi}, \emph{The Scalar-Tensor Theory of
  Gravitation}, Cambridge Monographs on Mathematical Physics. Cambridge
  University Press, 2003,
  \href{https://doi.org/10.1017/CBO9780511535093}{10.1017/CBO9780511535093}.

\bibitem{Clifton2011}
T.~Clifton, P.~G. Ferreira, A.~Padilla and C.~Skordis, \emph{Modified gravity
  and cosmology},
  \href{https://doi.org/https://doi.org/10.1016/j.physrep.2012.01.001}{\emph{Physics
  Reports} {\bfseries 513} (2012) 1}.

\bibitem{Joyce2014}
A.~Joyce, B.~Jain, J.~Khoury and M.~Trodden, \emph{{Beyond the Cosmological
  Standard Model}},
  \href{https://doi.org/10.1016/j.physrep.2014.12.002}{\emph{Phys. Rept.}
  {\bfseries 568} (2015) 1} [\href{https://arxiv.org/abs/1407.0059}{{\ttfamily
  1407.0059}}].

\bibitem{Dickey1994}
J.~O. Dickey, P.~L. Bender, J.~E. Faller, X.~X. Newhall, R.~L. Ricklefs, J.~G.
  Ries et~al., \emph{{Lunar Laser Ranging: A Continuing Legacy of the Apollo
  Program}}, \href{https://doi.org/10.1126/science.265.5171.482}{\emph{Science}
  {\bfseries 265} (1994) 482}.

\bibitem{Adelberger2003}
E.~Adelberger, B.~Heckel and A.~Nelson, \emph{{Tests of the Gravitational
  Inverse-Square Law}},
  \href{https://doi.org/10.1146/annurev.nucl.53.041002.110503}{\emph{Annual
  Review of Nuclear and Particle Science} {\bfseries 53} (2003) 77}.

\bibitem{Kapner2007}
D.~J. Kapner, T.~S. Cook, E.~G. Adelberger, J.~H. Gundlach, B.~R. Heckel, C.~D.
  Hoyle et~al., \emph{{Tests of the Gravitational Inverse-Square Law below the
  Dark-Energy Length Scale}},
  \href{https://doi.org/10.1103/PhysRevLett.98.021101}{\emph{Phys. Rev. Lett.}
  {\bfseries 98} (2007) 021101}.

\bibitem{Burrage:2016yjm}
C.~Burrage, E.~J. Copeland and P.~Millington, \emph{{Radial acceleration
  relation from symmetron fifth forces}},
  \href{https://doi.org/10.1103/PhysRevD.95.064050}{\emph{Phys. Rev. D}
  {\bfseries 95} (2017) 064050}
  [\href{https://arxiv.org/abs/1610.07529}{{\ttfamily 1610.07529}}].

\bibitem{OHare:2018ayv}
C.~A.~J. O'Hare and C.~Burrage, \emph{{Stellar kinematics from the symmetron
  fifth force in the Milky Way disk}},
  \href{https://doi.org/10.1103/PhysRevD.98.064019}{\emph{Phys. Rev. D}
  {\bfseries 98} (2018) 064019}
  [\href{https://arxiv.org/abs/1805.05226}{{\ttfamily 1805.05226}}].

\bibitem{Burrage:2018zuj}
C.~Burrage, E.~J. Copeland, C.~K\"ading and P.~Millington, \emph{{Symmetron
  scalar fields: Modified gravity, dark matter, or both?}},
  \href{https://doi.org/10.1103/PhysRevD.99.043539}{\emph{Phys. Rev. D}
  {\bfseries 99} (2019) 043539}
  [\href{https://arxiv.org/abs/1811.12301}{{\ttfamily 1811.12301}}].

\bibitem{Kading:2023hdb}
C.~K\"ading, \emph{{Lensing with Generalized Symmetrons}},
  \href{https://doi.org/10.3390/astronomy2020009}{\emph{Astronomy} {\bfseries
  2} (2023) 128} [\href{https://arxiv.org/abs/2304.05875}{{\ttfamily
  2304.05875}}].

\bibitem{Khoury2003}
J.~Khoury and A.~Weltman, \emph{{Chameleon cosmology}},
  \href{https://doi.org/10.1103/PhysRevD.69.044026}{\emph{Phys. Rev. D}
  {\bfseries 69} (2004) 044026}
  [\href{https://arxiv.org/abs/astro-ph/0309411}{{\ttfamily
  astro-ph/0309411}}].

\bibitem{Khoury20032}
J.~Khoury and A.~Weltman, \emph{{Chameleon fields: Awaiting surprises for tests
  of gravity in space}},
  \href{https://doi.org/10.1103/PhysRevLett.93.171104}{\emph{Phys. Rev. Lett.}
  {\bfseries 93} (2004) 171104}
  [\href{https://arxiv.org/abs/astro-ph/0309300}{{\ttfamily
  astro-ph/0309300}}].

\bibitem{Dehnen1992}
H.~Dehnen, H.~Frommert and F.~Ghaboussi, \emph{{Higgs field and a new scalar -
  tensor theory of gravity}},
  \href{https://doi.org/10.1007/BF00674344}{\emph{Int. J. Theor. Phys.}
  {\bfseries 31} (1992) 109}.

\bibitem{Gessner1992}
E.~Gessner, \emph{{A new scalar tensor theory for gravity and the flat rotation
  curves of spiral galaxies}},
  \href{https://doi.org/10.1007/BF00645239}{\emph{Astrophys. Space Sci.}
  {\bfseries 196} (1992) 29}.

\bibitem{Damour1994}
T.~Damour and A.~M. Polyakov, \emph{{The String dilaton and a least coupling
  principle}}, \href{https://doi.org/10.1016/0550-3213(94)90143-0}{\emph{Nucl.
  Phys. B} {\bfseries 423} (1994) 532}
  [\href{https://arxiv.org/abs/hep-th/9401069}{{\ttfamily hep-th/9401069}}].

\bibitem{Pietroni2005}
M.~Pietroni, \emph{Dark energy condensation},
  \href{https://doi.org/10.1103/PhysRevD.72.043535}{\emph{Phys. Rev. D}
  {\bfseries 72} (2005) 043535}.

\bibitem{Olive2008}
K.~A. Olive and M.~Pospelov, \emph{Environmental dependence of masses and
  coupling constants},
  \href{https://doi.org/10.1103/PhysRevD.77.043524}{\emph{Phys. Rev. D}
  {\bfseries 77} (2008) 043524}.

\bibitem{Brax2010}
P.~Brax, C.~van~de Bruck, A.-C. Davis and D.~Shaw, \emph{Dilaton and modified
  gravity}, \href{https://doi.org/10.1103/PhysRevD.82.063519}{\emph{Phys. Rev.
  D} {\bfseries 82} (2010) 063519}.

\bibitem{Hinterbichler2010}
K.~Hinterbichler and J.~Khoury, \emph{{Symmetron Fields: Screening Long-Range
  Forces Through Local Symmetry Restoration}},
  \href{https://doi.org/10.1103/PhysRevLett.104.231301}{\emph{Phys. Rev. Lett.}
  {\bfseries 104} (2010) 231301}
  [\href{https://arxiv.org/abs/1001.4525}{{\ttfamily 1001.4525}}].

\bibitem{Hinterbichler2011}
K.~Hinterbichler, J.~Khoury, A.~Levy and A.~Matas, \emph{{Symmetron
  Cosmology}}, \href{https://doi.org/10.1103/PhysRevD.84.103521}{\emph{Phys.
  Rev. D} {\bfseries 84} (2011) 103521}
  [\href{https://arxiv.org/abs/1107.2112}{{\ttfamily 1107.2112}}].

\bibitem{Gasperini:2001pc}
M.~Gasperini, F.~Piazza and G.~Veneziano, \emph{{Quintessence as a runaway
  dilaton}}, \href{https://doi.org/10.1103/PhysRevD.65.023508}{\emph{Phys. Rev.
  D} {\bfseries 65} (2002) 023508}
  [\href{https://arxiv.org/abs/gr-qc/0108016}{{\ttfamily gr-qc/0108016}}].

\bibitem{Damour:2002nv}
T.~Damour, F.~Piazza and G.~Veneziano, \emph{{Violations of the equivalence
  principle in a dilaton runaway scenario}},
  \href{https://doi.org/10.1103/PhysRevD.66.046007}{\emph{Phys. Rev. D}
  {\bfseries 66} (2002) 046007}
  [\href{https://arxiv.org/abs/hep-th/0205111}{{\ttfamily hep-th/0205111}}].

\bibitem{Damour:2002mi}
T.~Damour, F.~Piazza and G.~Veneziano, \emph{{Runaway dilaton and equivalence
  principle violations}},
  \href{https://doi.org/10.1103/PhysRevLett.89.081601}{\emph{Phys. Rev. Lett.}
  {\bfseries 89} (2002) 081601}
  [\href{https://arxiv.org/abs/gr-qc/0204094}{{\ttfamily gr-qc/0204094}}].

\bibitem{Brax:2010gi}
P.~Brax, C.~van~de Bruck, A.-C. Davis and D.~Shaw, \emph{{The Dilaton and
  Modified Gravity}},
  \href{https://doi.org/10.1103/PhysRevD.82.063519}{\emph{Phys. Rev. D}
  {\bfseries 82} (2010) 063519}
  [\href{https://arxiv.org/abs/1005.3735}{{\ttfamily 1005.3735}}].

\bibitem{Brax:2011ja}
P.~Brax, C.~van~de Bruck, A.-C. Davis, B.~Li and D.~J. Shaw, \emph{{Nonlinear
  Structure Formation with the Environmentally Dependent Dilaton}},
  \href{https://doi.org/10.1103/PhysRevD.83.104026}{\emph{Phys. Rev. D}
  {\bfseries 83} (2011) 104026}
  [\href{https://arxiv.org/abs/1102.3692}{{\ttfamily 1102.3692}}].

\bibitem{Brax2022}
P.~Brax, H.~Fischer, C.~K\"ading and M.~Pitschmann, \emph{{The environment
  dependent dilaton in the laboratory and the solar system}},
  \href{https://doi.org/10.1140/epjc/s10052-022-10905-w}{\emph{Eur. Phys. J. C}
  {\bfseries 82} (2022) 934}
  [\href{https://arxiv.org/abs/2203.12512}{{\ttfamily 2203.12512}}].

\bibitem{Dvali2000}
G.~R. Dvali, G.~Gabadadze and M.~Porrati, \emph{{4-D gravity on a brane in 5-D
  Minkowski space}},
  \href{https://doi.org/10.1016/S0370-2693(00)00669-9}{\emph{Phys. Lett. B}
  {\bfseries 485} (2000) 208}
  [\href{https://arxiv.org/abs/hep-th/0005016}{{\ttfamily hep-th/0005016}}].

\bibitem{Nicolis2008}
A.~Nicolis, R.~Rattazzi and E.~Trincherini, \emph{{The Galileon as a local
  modification of gravity}},
  \href{https://doi.org/10.1103/PhysRevD.79.064036}{\emph{Phys. Rev. D}
  {\bfseries 79} (2009) 064036}
  [\href{https://arxiv.org/abs/0811.2197}{{\ttfamily 0811.2197}}].

\bibitem{Ali2012}
A.~Ali, R.~Gannouji, M.~W. Hossain and M.~Sami, \emph{{Light mass galileons:
  Cosmological dynamics, mass screening and observational constraints}},
  \href{https://doi.org/10.1016/j.physletb.2012.10.009}{\emph{Phys. Lett. B}
  {\bfseries 718} (2012) 5} [\href{https://arxiv.org/abs/1207.3959}{{\ttfamily
  1207.3959}}].

\bibitem{Burrage:2016bwy}
C.~Burrage and J.~Sakstein, \emph{{A Compendium of Chameleon Constraints}},
  \href{https://doi.org/10.1088/1475-7516/2016/11/045}{\emph{JCAP} {\bfseries
  11} (2016) 045} [\href{https://arxiv.org/abs/1609.01192}{{\ttfamily
  1609.01192}}].

\bibitem{Burrage:2017qrf}
C.~Burrage and J.~Sakstein, \emph{{Tests of Chameleon Gravity}},
  \href{https://doi.org/10.1007/s41114-018-0011-x}{\emph{Living Rev. Rel.}
  {\bfseries 21} (2018) 1} [\href{https://arxiv.org/abs/1709.09071}{{\ttfamily
  1709.09071}}].

\bibitem{Pokotilovski:2012xuk}
Y.~N. Pokotilovski, \emph{{Strongly coupled chameleon fields: Possible test
  with a neutron Lloyd's mirror interferometer}},
  \href{https://doi.org/10.1016/j.physletb.2013.01.022}{\emph{Phys. Lett. B}
  {\bfseries 719} (2013) 341}
  [\href{https://arxiv.org/abs/1203.5017}{{\ttfamily 1203.5017}}].

\bibitem{Pokotilovski:2013tma}
Y.~N. Pokotilovski, \emph{{Potential of the neutron Lloyd`s mirror
  interferometer for the search for new interactions}},
  \href{https://doi.org/10.1134/S106377611309001X}{\emph{J. Exp. Theor. Phys.}
  {\bfseries 116} (2013) 609}
  [\href{https://arxiv.org/abs/1311.4679}{{\ttfamily 1311.4679}}].

\bibitem{Jenke:2014yel}
T.~Jenke et~al., \emph{{Gravity Resonance Spectroscopy Constrains Dark Energy
  and Dark Matter Scenarios}},
  \href{https://doi.org/10.1103/PhysRevLett.112.151105}{\emph{Phys. Rev. Lett.}
  {\bfseries 112} (2014) 151105}
  [\href{https://arxiv.org/abs/1404.4099}{{\ttfamily 1404.4099}}].

\bibitem{Baum:2014rka}
S.~Baum, G.~Cantatore, D.~H.~H. Hoffmann, M.~Karuza, Y.~K. Semertzidis,
  A.~Upadhye et~al., \emph{{Detecting solar chameleons through radiation
  pressure}}, \href{https://doi.org/10.1016/j.physletb.2014.10.055}{\emph{Phys.
  Lett. B} {\bfseries 739} (2014) 167}
  [\href{https://arxiv.org/abs/1409.3852}{{\ttfamily 1409.3852}}].

\bibitem{Burrage:2014oza}
C.~Burrage, E.~J. Copeland and E.~A. Hinds, \emph{{Probing Dark Energy with
  Atom Interferometry}},
  \href{https://doi.org/10.1088/1475-7516/2015/03/042}{\emph{JCAP} {\bfseries
  03} (2015) 042} [\href{https://arxiv.org/abs/1408.1409}{{\ttfamily
  1408.1409}}].

\bibitem{Hamilton:2015zga}
P.~Hamilton, M.~Jaffe, P.~Haslinger, Q.~Simmons, H.~M\"uller and J.~Khoury,
  \emph{{Atom-interferometry constraints on dark energy}},
  \href{https://doi.org/10.1126/science.aaa8883}{\emph{Science} {\bfseries 349}
  (2015) 849} [\href{https://arxiv.org/abs/1502.03888}{{\ttfamily
  1502.03888}}].

\bibitem{Lemmel:2015kwa}
H.~Lemmel, P.~Brax, A.~N. Ivanov, T.~Jenke, G.~Pignol, M.~Pitschmann et~al.,
  \emph{{Neutron Interferometry constrains dark energy chameleon fields}},
  \href{https://doi.org/10.1016/j.physletb.2015.02.063}{\emph{Phys. Lett. B}
  {\bfseries 743} (2015) 310}
  [\href{https://arxiv.org/abs/1502.06023}{{\ttfamily 1502.06023}}].

\bibitem{Burrage:2015lya}
C.~Burrage and E.~J. Copeland, \emph{{Using Atom Interferometry to Detect Dark
  Energy}}, \href{https://doi.org/10.1080/00107514.2015.1060058}{\emph{Contemp.
  Phys.} {\bfseries 57} (2016) 164}
  [\href{https://arxiv.org/abs/1507.07493}{{\ttfamily 1507.07493}}].

\bibitem{Elder:2016yxm}
B.~Elder, J.~Khoury, P.~Haslinger, M.~Jaffe, H.~M\"uller and P.~Hamilton,
  \emph{{Chameleon Dark Energy and Atom Interferometry}},
  \href{https://doi.org/10.1103/PhysRevD.94.044051}{\emph{Phys. Rev. D}
  {\bfseries 94} (2016) 044051}
  [\href{https://arxiv.org/abs/1603.06587}{{\ttfamily 1603.06587}}].

\bibitem{Ivanov:2016rfs}
A.~N. Ivanov, G.~Cronenberg, R.~H\"ollwieser, M.~Pitschmann, T.~Jenke,
  M.~Wellenzohn et~al., \emph{{Exact solution for chameleon field, self-coupled
  through the Ratra-Peebles potential with $n=1$ and confined between two
  parallel plates}},
  \href{https://doi.org/10.1103/PhysRevD.94.085005}{\emph{Phys. Rev. D}
  {\bfseries 94} (2016) 085005}
  [\href{https://arxiv.org/abs/1606.06867}{{\ttfamily 1606.06867}}].

\bibitem{Burrage:2016rkv}
C.~Burrage, A.~Kuribayashi-Coleman, J.~Stevenson and B.~Thrussell,
  \emph{{Constraining symmetron fields with atom interferometry}},
  \href{https://doi.org/10.1088/1475-7516/2016/12/041}{\emph{JCAP} {\bfseries
  12} (2016) 041} [\href{https://arxiv.org/abs/1609.09275}{{\ttfamily
  1609.09275}}].

\bibitem{Jaffe:2016fsh}
M.~Jaffe, P.~Haslinger, V.~Xu, P.~Hamilton, A.~Upadhye, B.~Elder et~al.,
  \emph{{Author Correction: Testing sub-gravitational forces on atoms from a
  miniature in-vacuum source mass [doi: 10.1038/nphys4189]}},
  \href{https://doi.org/10.1038/s41567-023-02255-5}{\emph{Nature Phys.}
  {\bfseries 13} (2017) 938}
  [\href{https://arxiv.org/abs/1612.05171}{{\ttfamily 1612.05171}}].

\bibitem{Brax:2017hna}
P.~Brax and M.~Pitschmann, \emph{{Exact solutions to nonlinear symmetron
  theory: One- and two-mirror systems}},
  \href{https://doi.org/10.1103/PhysRevD.97.064015}{\emph{Phys. Rev. D}
  {\bfseries 97} (2018) 064015}
  [\href{https://arxiv.org/abs/1712.09852}{{\ttfamily 1712.09852}}].

\bibitem{Sabulsky:2018jma}
D.~O. Sabulsky, I.~Dutta, E.~A. Hinds, B.~Elder, C.~Burrage and E.~J. Copeland,
  \emph{{Experiment to detect dark energy forces using atom interferometry}},
  \href{https://doi.org/10.1103/PhysRevLett.123.061102}{\emph{Phys. Rev. Lett.}
  {\bfseries 123} (2019) 061102}
  [\href{https://arxiv.org/abs/1812.08244}{{\ttfamily 1812.08244}}].

\bibitem{Brax:2018iyo}
P.~Brax, C.~Burrage and A.-C. Davis, \emph{{Laboratory constraints}},
  \href{https://doi.org/10.1142/S0218271818480097}{\emph{Int. J. Mod. Phys. D}
  {\bfseries 27} (2018) 1848009}.

\bibitem{Cronenberg:2018qxf}
G.~Cronenberg, P.~Brax, H.~Filter, P.~Geltenbort, T.~Jenke, G.~Pignol et~al.,
  \emph{{Acoustic Rabi oscillations between gravitational quantum states and
  impact on symmetron dark energy}},
  \href{https://doi.org/10.1038/s41567-018-0205-x}{\emph{Nature Phys.}
  {\bfseries 14} (2018) 1022}
  [\href{https://arxiv.org/abs/1902.08775}{{\ttfamily 1902.08775}}].

\bibitem{Zhang:2019hos}
X.~Zhang, R.~Niu and W.~Zhao, \emph{{Constraining the scalar-tensor gravity
  theories with and without screening mechanisms by combined observations}},
  \href{https://doi.org/10.1103/PhysRevD.100.024038}{\emph{Phys. Rev. D}
  {\bfseries 100} (2019) 024038}
  [\href{https://arxiv.org/abs/1906.10791}{{\ttfamily 1906.10791}}].

\bibitem{ArguedasCuendis:2019fxj}
S.~Arguedas~Cuendis et~al., \emph{{First Results on the Search for Chameleons
  with the KWISP Detector at CAST}},
  \href{https://doi.org/10.1016/j.dark.2019.100367}{\emph{Phys. Dark Univ.}
  {\bfseries 26} (2019) 100367}
  [\href{https://arxiv.org/abs/1906.01084}{{\ttfamily 1906.01084}}].

\bibitem{Jenke:2020obe}
T.~Jenke, J.~Bosina, J.~Micko, M.~Pitschmann, R.~Sedmik and H.~Abele,
  \emph{{Gravity resonance spectroscopy and dark energy symmetron fields:
  qBOUNCE experiments performed with Rabi and Ramsey spectroscopy}},
  \href{https://doi.org/10.1140/epjs/s11734-021-00088-y}{\emph{Eur. Phys. J.
  ST} {\bfseries 230} (2021) 1131}
  [\href{https://arxiv.org/abs/2012.07472}{{\ttfamily 2012.07472}}].

\bibitem{Pitschmann:2020ejb}
M.~Pitschmann, \emph{{Exact solutions to nonlinear symmetron theory: One- and
  two-mirror systems. II.}},
  \href{https://doi.org/10.1103/PhysRevD.103.084013}{\emph{Phys. Rev. D}
  {\bfseries 103} (2021) 084013}
  [\href{https://arxiv.org/abs/2012.12752}{{\ttfamily 2012.12752}}].

\bibitem{Brax:2021wcv}
P.~Brax, S.~Casas, H.~Desmond and B.~Elder, \emph{{Testing Screened Modified
  Gravity}}, \href{https://doi.org/10.3390/universe8010011}{\emph{Universe}
  {\bfseries 8} (2021) 11} [\href{https://arxiv.org/abs/2201.10817}{{\ttfamily
  2201.10817}}].

\bibitem{Qvarfort:2021zrl}
S.~Qvarfort, D.~R\"atzel and S.~Stopyra, \emph{{Constraining modified gravity
  with quantum optomechanics}},
  \href{https://doi.org/10.1088/1367-2630/ac3e1b}{\emph{New J. Phys.}
  {\bfseries 24} (2022) 033009}
  [\href{https://arxiv.org/abs/2108.00742}{{\ttfamily 2108.00742}}].

\bibitem{Yin:2022geb}
P.~Yin, R.~Li, C.~Yin, X.~Xu, X.~Bian, H.~Xie et~al., \emph{{Experiments with
  levitated force sensor challenge theories of dark energy}},
  \href{https://doi.org/10.1038/s41567-022-01706-9}{\emph{Nature Phys.}
  {\bfseries 18} (2022) 1181}.

\bibitem{Betz:2022djh}
J.~Betz, J.~Manley, E.~M. Wright, D.~Grin and S.~Singh, \emph{{Searching for
  Chameleon Dark Energy with Mechanical Systems}},
  \href{https://doi.org/10.1103/PhysRevLett.129.131302}{\emph{Phys. Rev. Lett.}
  {\bfseries 129} (2022) 131302}
  [\href{https://arxiv.org/abs/2201.12372}{{\ttfamily 2201.12372}}].

\bibitem{Brax:2022olf}
P.~Brax, A.-C. Davis and B.~Elder, \emph{{Screened scalar fields in hydrogen
  and muonium}}, \href{https://doi.org/10.1103/PhysRevD.107.044008}{\emph{Phys.
  Rev. D} {\bfseries 107} (2023) 044008}
  [\href{https://arxiv.org/abs/2207.11633}{{\ttfamily 2207.11633}}].

\bibitem{Hartley:2019wzu}
D.~Hartley, C.~K\"ading, R.~Howl and I.~Fuentes, \emph{{Quantum-enhanced
  screened dark energy detection}},
  \href{https://doi.org/10.1140/epjc/s10052-023-12360-7}{\emph{Eur. Phys. J. C}
  {\bfseries 84} (2024) 49} [\href{https://arxiv.org/abs/1909.02272}{{\ttfamily
  1909.02272}}].

\bibitem{Fischer:2023koa}
H.~Fischer, C.~K\"ading, R.~I.~P. Sedmik, H.~Abele, P.~Brax and M.~Pitschmann,
  \emph{{Search for environment-dependent dilatons}},
  \href{https://doi.org/10.1016/j.dark.2024.101419}{\emph{Phys. Dark Univ.}
  {\bfseries 43} (2024) 101419}
  [\href{https://arxiv.org/abs/2307.00243}{{\ttfamily 2307.00243}}].

\bibitem{Fischer:2023eww}
H.~Fischer, C.~K\"ading, H.~Lemmel, S.~Sponar and M.~Pitschmann, \emph{{Search
  for dark energy with neutron interferometry}},
  \href{https://doi.org/10.1093/ptep/ptae014}{\emph{PTEP} {\bfseries 2024}
  (2024) 023E02} [\href{https://arxiv.org/abs/2310.18109}{{\ttfamily
  2310.18109}}].

\bibitem{Fischer:2024coj}
H.~Fischer and R.~I.~P. Sedmik, \emph{{Numerical Methods for Scalar Field Dark
  Energy in Table-top Experiments and Lunar Laser Ranging}},
  \href{https://arxiv.org/abs/2401.16179}{{\ttfamily 2401.16179}}.

\bibitem{Klimchitskaya:2024dvk}
G.~L. Klimchitskaya and V.~M. Mostepanenko, \emph{{The Nature of Dark Energy
  and Constraints on Its Hypothetical Constituents from Force Measurements}},
  \href{https://doi.org/10.3390/universe10030119}{\emph{Universe} {\bfseries
  10} (2024) 119} [\href{https://arxiv.org/abs/2403.05988}{{\ttfamily
  2403.05988}}].

\bibitem{Rene}
H.~Haghmoradi, H.~Fischer, A.~Bertolini, I.~Gali\'c, F.~Intravaia,
  M.~Pitschmann et~al., \emph{{Force metrology with plane parallel plates:
  Final design review and outlook}}, {\emph{MDPI Physics} {\bfseries 6} (2024)
  } [\href{https://arxiv.org/abs/2403.10998}{{\ttfamily 2403.10998}}].

\bibitem{Brax:2018grq}
P.~Brax and S.~Fichet, \emph{{Quantum Chameleons}},
  \href{https://doi.org/10.1103/PhysRevD.99.104049}{\emph{Phys. Rev. D}
  {\bfseries 99} (2019) 104049}
  [\href{https://arxiv.org/abs/1809.10166}{{\ttfamily 1809.10166}}].

\bibitem{Burrage:2018pyg}
C.~Burrage, C.~K\"ading, P.~Millington and J.~Min\'a\v{r}, \emph{{Open quantum
  dynamics induced by light scalar fields}},
  \href{https://doi.org/10.1103/PhysRevD.100.076003}{\emph{Phys. Rev. D}
  {\bfseries 100} (2019) 076003}
  [\href{https://arxiv.org/abs/1812.08760}{{\ttfamily 1812.08760}}].

\bibitem{Burrage:2019szw}
C.~Burrage, C.~K\"ading, P.~Millington and J.~Min\'a\v{r}, \emph{{Influence
  functionals, decoherence and conformally coupled scalars}},
  \href{https://doi.org/10.1088/1742-6596/1275/1/012041}{\emph{J. Phys. Conf.
  Ser.} {\bfseries 1275} (2019) 012041}
  [\href{https://arxiv.org/abs/1902.09607}{{\ttfamily 1902.09607}}].

\bibitem{Kading:2023mdk}
C.~K\"ading, M.~Pitschmann and C.~Voith, \emph{{Dilaton-induced open quantum
  dynamics}}, \href{https://doi.org/10.1140/epjc/s10052-023-11939-4}{\emph{Eur.
  Phys. J. C} {\bfseries 83} (2023) 767}
  [\href{https://arxiv.org/abs/2306.10896}{{\ttfamily 2306.10896}}].

\bibitem{Baez-Camargo:2024jia}
A.~L. B\'aez-Camargo, D.~Hartley, C.~K\"ading and I.~Fuentes-Guridi,
  \emph{{Dynamical Casimir effect with screened scalar fields}},
  \href{https://arxiv.org/abs/2404.02630}{{\ttfamily 2404.02630}}.

\bibitem{Hartley:2018lvm}
D.~Hartley, C.~K\"ading, R.~Howl and I.~Fuentes, \emph{{Quantum simulation of
  dark energy candidates}},
  \href{https://doi.org/10.1103/PhysRevD.99.105002}{\emph{Phys. Rev. D}
  {\bfseries 99} (2019) 105002}
  [\href{https://arxiv.org/abs/1811.06927}{{\ttfamily 1811.06927}}].

\bibitem{Sedmik:2021iaw}
R.~I.~P. Sedmik and M.~Pitschmann, \emph{{Next Generation Design and Prospects
  for Cannex}}, \href{https://doi.org/10.3390/universe7070234}{\emph{Universe}
  {\bfseries 7} (2021) 234} [\href{https://arxiv.org/abs/2107.07645}{{\ttfamily
  2107.07645}}].

\bibitem{Brax:2013cfa}
P.~Brax, G.~Pignol and D.~Roulier, \emph{{Probing Strongly Coupled Chameleons
  with Slow Neutrons}},
  \href{https://doi.org/10.1103/PhysRevD.88.083004}{\emph{Phys. Rev. D}
  {\bfseries 88} (2013) 083004}
  [\href{https://arxiv.org/abs/1306.6536}{{\ttfamily 1306.6536}}].

\bibitem{VAINSHTEIN1972393}
A.~Vainshtein, \emph{To the problem of nonvanishing gravitation mass},
  \href{https://doi.org/https://doi.org/10.1016/0370-2693(72)90147-5}{\emph{Physics
  Letters B} {\bfseries 39} (1972) 393}.

\bibitem{PhysRevD.81.065019}
H.~Abele, T.~Jenke, H.~Leeb and J.~Schmiedmayer, \emph{Ramsey's method of
  separated oscillating fields and its application to gravitationally induced
  quantum phase shifts},
  \href{https://doi.org/10.1103/PhysRevD.81.065019}{\emph{Phys. Rev. D}
  {\bfseries 81} (2010) 065019}.

\bibitem{Jenke:2011zz}
T.~Jenke, P.~Geltenbort, H.~Lemmel and H.~Abele, \emph{{Realization of a
  gravity-resonance-spectroscopy technique}},
  \href{https://doi.org/10.1038/nphys1970}{\emph{Nature Phys.} {\bfseries 7}
  (2011) 468}.

\bibitem{Sponar:2020gfr}
S.~Sponar, R.~I.~P. Sedmik, M.~Pitschmann, H.~Abele and Y.~Hasegawa,
  \emph{{Tests of fundamental quantum mechanics and dark interactions with
  low-energy neutrons}},
  \href{https://doi.org/10.1038/s42254-021-00298-2}{\emph{Nature Rev. Phys.}
  {\bfseries 3} (2021) 309} [\href{https://arxiv.org/abs/2012.09048}{{\ttfamily
  2012.09048}}].

\bibitem{Rauch74}
H.~Rauch, W.~Treimer and U.~Bonse, \emph{Test of a single crystal neutron
  interferometer}, \href{https://doi.org/DOI:
  10.1016/0375-9601(74)90132-7}{\emph{Phys. Lett.} {\bfseries 47A} (1974) 369
  }.

\bibitem{RauchBook}
H.~Rauch and S.~A. Werner, \emph{Neutron Interferometry}. Clarendon Press,
  Oxford, 2000.

\bibitem{muller2019lunar}
J.~M{\"u}ller, T.~W. Murphy, U.~Schreiber, P.~J. Shelus, J.-M. Torre, J.~G.
  Williams et~al., \emph{Lunar laser ranging: a tool for general relativity,
  lunar geophysics and earth science},
  \href{https://doi.org/10.1007/s00190-019-01296-0}{\emph{J. Geodesy}
  {\bfseries 93} (2019) 2195}.

\bibitem{Sakstein:2017pqi}
J.~Sakstein, \emph{{Tests of Gravity with Future Space-Based Experiments}},
  \href{https://doi.org/10.1103/PhysRevD.97.064028}{\emph{Phys. Rev. D}
  {\bfseries 97} (2018) 064028}
  [\href{https://arxiv.org/abs/1710.03156}{{\ttfamily 1710.03156}}].

\bibitem{Kraiselburd:2015vyf}
L.~Kraiselburd, S.~J. Landau, M.~Salgado, D.~Sudarsky and H.~Vucetich,
  \emph{{Equivalence Principle in Chameleon Models}},
  \href{https://doi.org/10.1103/PhysRevD.97.104044}{\emph{Phys. Rev. D}
  {\bfseries 97} (2018) 104044}
  [\href{https://arxiv.org/abs/1511.06307}{{\ttfamily 1511.06307}}].

\bibitem{Hofmann:2018myc}
F.~Hofmann and J.~M\"uller, \emph{{Relativistic tests with lunar laser
  ranging}}, \href{https://doi.org/10.1088/1361-6382/aa8f7a}{\emph{Class.
  Quant. Grav.} {\bfseries 35} (2018) 035015}.

\bibitem{MarioHabil}
M.~Pitschmann, \emph{{The High Precision Frontier: Search for New Physics with
  “Tabletop Experiments” \& Beyond}}, habilitation, TU Wien, 2023.

\bibitem{Adelberger:2003zx}
E.~G. Adelberger, B.~R. Heckel and A.~E. Nelson, \emph{{Tests of the
  gravitational inverse square law}},
  \href{https://doi.org/10.1146/annurev.nucl.53.041002.110503}{\emph{Ann. Rev.
  Nucl. Part. Sci.} {\bfseries 53} (2003) 77}
  [\href{https://arxiv.org/abs/hep-ph/0307284}{{\ttfamily hep-ph/0307284}}].

\bibitem{Upadhye:2012rc}
A.~Upadhye, \emph{{Symmetron dark energy in laboratory experiments}},
  \href{https://doi.org/10.1103/PhysRevLett.110.031301}{\emph{Phys. Rev. Lett.}
  {\bfseries 110} (2013) 031301}
  [\href{https://arxiv.org/abs/1210.7804}{{\ttfamily 1210.7804}}].

\end{thebibliography}\endgroup
\bibliographystyle{JHEP}


\end{document}